\documentclass[transmag]{IEEEtran}
\usepackage{latexsym}
\usepackage{graphicx}
\usepackage{amsfonts,amssymb,amsmath}
\usepackage{hyperref}
\usepackage{booktabs}
\usepackage[inline]{enumitem}
\usepackage[export]{adjustbox}
\usepackage{wrapfig}
\usepackage[caption=false,font=footnotesize]{subfig}
\usepackage{amsfonts} 
\usepackage{amsthm} 
\usepackage{commath} 
\usepackage{algorithmic}
\usepackage{mathtools, nccmath}

\usepackage{comment} 
\usepackage{cite}
\usepackage{array}
\usepackage{textcomp}
\usepackage{xcolor}
\usepackage[final]{pdfpages}

\def\BibTeX{{\rm B\kern-.05em{\sc i\kern-.025em b}\kern-.08em T\kern-.1667em\lower.7ex\hbox{E}\kern-.125emX}}

\newcommand{\red}{\protect\textcolor{red}}

\begin{document}

\title{Comparing Power Processing System Approaches\\ in Second-Use Battery Energy Buffering \\ for Electric Vehicle Charging}
\author{Xiaofan~Cui\IEEEauthorrefmark{1}\thanks{\IEEEauthorrefmark{1} The authors are with the Department of Electrical Engineering and Computer Science, University of Michigan, Ann Arbor, MI, 48109, USA (e-mail: cuixf@umich.edu (corresponding author); aramyar@umich.edu; avestruz@umich.edu).}\!\!,~\IEEEmembership{Student Member,~IEEE,} 
Alireza~Ramyar\IEEEauthorrefmark{1}\!\!,~\IEEEmembership{Student Member,~IEEE,}  \\ 
Jason~B.~Siegel\IEEEauthorrefmark{2} \thanks{\IEEEauthorrefmark{2} The authors are with the Department of Mechanical Engineering, University of Michigan, Ann Arbor, MI, 48109, USA (e-mail: siegeljb@umich.edu; pmohtat@umich.edu;  annastef@umich.edu).}\!\!\!\!,~\IEEEmembership{Senior Member,~IEEE,}
Peyman~Mohtat\IEEEauthorrefmark{2}\!\!,~\IEEEmembership{Student Member,~IEEE,}\\
Anna~G.~Stefanopoulou\IEEEauthorrefmark{2}\!\!,~\IEEEmembership{Fellow,~IEEE,}
and~Al-Thaddeus~Avestruz\IEEEauthorrefmark{1}\!\!,~\IEEEmembership{Member,~IEEE}
\thanks{This work was supported in part by the Michigan Transportation Research and Commercialization (MTRAC) Grant CASE-283536 of the 21st Century Jobs Trust Fund received through the MSF from the State of Michigan.}
\thanks{This manuscript was accepted by Journal of Energy Storage on January 8, 2022.}}

\IEEEtitleabstractindextext{
\begin{abstract}
The heterogeneity in pack voltages and capacity of aged packs limits the performance and economic viability of second-use battery energy storage systems (2-BESS) due to issues of reliability and available energy. Overcoming these limitations could enable extended use of batteries and improve environmental impacts of electric vehicles by reducing the number of batteries produced. This paper compares Lite-Sparse Hierarchical Partial Power Processing (LS-HiPPP), a new method for power processing in 2-BESS, to conventional power processing architectures using a stochastic EV charging plaza model. This method for performance evaluation allows a fair comparison among power processing architectures for 2-BESS.
Results show that LS-HiPPP increases the battery energy utilization to 94\% as compared to 78\% for conventional partial power processing (C-PPP) and 23\% for full power processing. These results were obtained with 25\%  heterogeneity in individual battery capacities and 20\% power processing within the 2-BESS. Derating and captured value are two derived performance metrics for comparing LS-HiPPP and C-PPP in this work. The derating for LS-HiPPP is 84.3\% in comparison to 63.1\% for C-PPP. The captured value for LS-HiPPP is 79.8\% versus 51\% for C-PPP.

\end{abstract}
}

\maketitle

\section{INTRODUCTION} \label{sec:Intro}
\IEEEPARstart{A}{s} the number and power levels of electric vehicle chargers increase so will the stress on the electric grid \cite{Deb2018}. Energy buffering, consisting of point of use energy storage, smooths peak power stress on the grid while supplying EV charging demand \cite{Bryden2019}. Additionally, energy buffering reduces the capital expense of grid upgrades in commissioning EV chargers \cite{Bryden2019,DArpino2019} while reducing the cost from utility tariffs associated with peak demands \cite{Yang2019}.

\begin{figure*}[ht]
    \centering
    \includegraphics[width=5in]{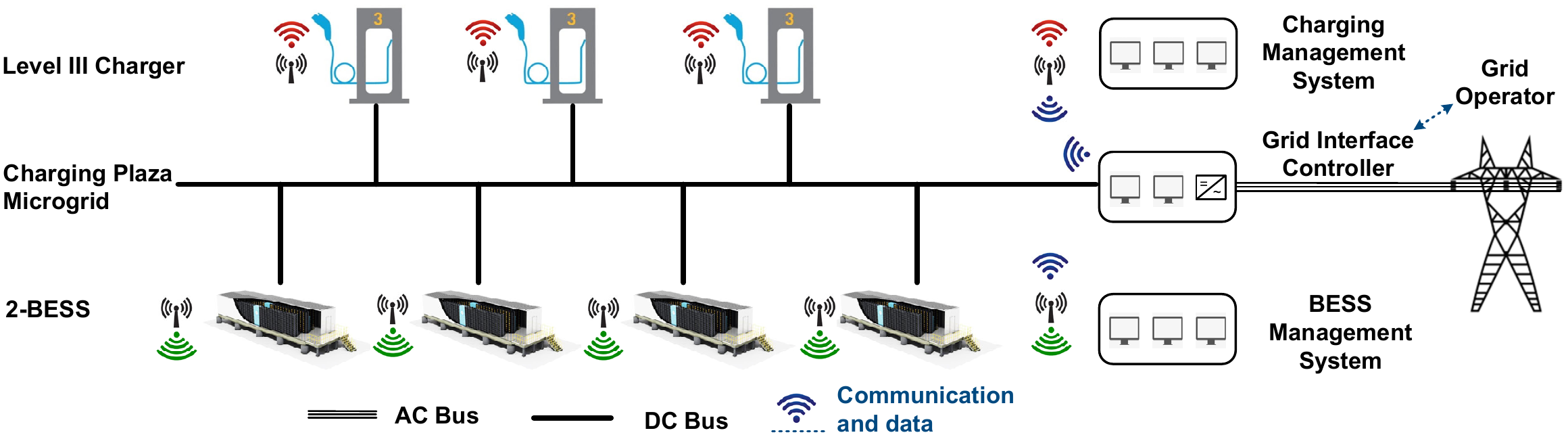}
    \caption{\label{fig:new-charging-plaza} An EV Charging Plaza is an example of an application for 2-BESS to reduced the peak grid power demand. The plaza contains a DC microgrid with multiple EV chargers, battery energy storage systems, and a single connection to grid. A simplified charging model is used to compare the performance of the Lite-Sparse Hierarchical Partial Power Processing (LS-HiPPP) and conventional power processing architectures for the 2-BESS.}
\end{figure*}

The second-use of batteries could be a sustainable outlet for the influx of used batteries that will accompany the proliferation of electric vehicles \cite{Casals2019}. The goal of this research is to enable second-use EV batteries as a cost-competitive solution for energy storage.  Several economic obstacles can be addressed by optimizing the power electronics architecture for second-use battery energy storage systems (2-BESS) including the cost of power electronics, transportation and inventory of used batteries \cite{Slattery2021}, and the cost of testing \cite{Neubauer2015}. Transportation costs can be reduced by locally supplying the 2-BESS production 
centers.
The inventory costs can be reduced by just-in-time production. 
Just-in-time battery production using the purely local supply causes significant heterogeneity in the battery capacities because no large inventory can be utilized to sort, group and homogenize the batteries. This significant battery heterogeneity needs to be adapted by the new power processing architecture.
Moreover, the cost of testing batteries prior to assembly of the 2-BESS can be reduced if heterogeneity in the battery capacities can be accommodated by the power processing architecture and design.

We consider an EV charging plaza illustrated in Fig.\,\ref{fig:new-charging-plaza}, as a case-study for the feasibility of 2-BESS deployment. The plaza consists of one grid connection with several EV chargers that are connected to multiple energy storage systems through a dc microgrid \cite{Leonori2020,plazaDutch2019} and regulated by three coordinated controllers. The stress on the grid from EV charging is well-known, making stationary energy storage an important mitigation  \cite{Huda2020} \cite{Pearre2016}. Grid operators can impose demand response restrictions by communicating both pricing and hard power constraints in real time, which are used to determine how best to allocate the energy stored in the battery \cite{yao2017, Xiong2018}. Balancing the demand for EV charging with the grid supply together with the size of the storage for energy buffering can be challenging making accurate forecasting valuable. A data-driven approach to anticipate EV charging demand from real-world traffic data in China has been proposed \cite{Xing2019}.  Another forecasting approach uses historical traffic and weather data from South Korea to formulate a model of EV demand \cite{Arias2016}. Large scale EV charging together with BESS operation can be coordinated by a central controller through a communications network to maintain grid stability \cite{Lee2016}.


Other important applications of battery energy storage that could benefit from optimized power electronics architectures for 2-BESS include supporting integration of renewable energy generation. 
Collocating the battery energy storage systems with solar \cite{Stroe2018} and wind power \cite{Braun2011} has been particularly successful in reducing variability in renewable power generation in Denmark.

Accurate forecasting models of intermittent generation, storage capacity, and demand are necessary for optimizing the utilization of renewable energy and EV charging applications.  Collocating stationary energy storage with photovoltaic plants dramatically increases the energy utilization of the generation \cite{Mazzeo2021}.  We expect similar benefits of collocating stationary energy storage with EV charging \cite{Mazzeo2015}.

Cost and reliability are the two most important factors for selecting battery chemistry and power electronics architectures for energy storage systems. The feasibility of using 2-BESS for grid support has been successfully demonstrated in \cite{Aziz2015}.  In Denmark, 2-BESS together with full power processing (FPP) was used for residential demand response for the grid to ensure the reliability of the system \cite{Saez-De-Ibarra2015}. Although second-use batteries for BESS have lower upfront costs relative to new batteries, current 2-BESS system architectures based on FPP may have higher levelized cost of storage \cite{Steckel2021}, which underscores the need for innovation.  Government policy can jumpstart the innovation needed for expansion of 2-BESS by providing economic incentives for choosing sustainable technology solutions \cite{Gu2021}. The second-use of automotive batteries improves their sustainability by reducing the bottleneck in recycling \cite{Abdelbaky2021} and enables the higher utilization of intermittent renewable energy.




The benchmark architecture for 2-BESS is Full Power Processing (FPP), shown in Fig.~\ref{fig:Topo_Comp}(a), which accommodates the variation in battery capacity and obtains high reliability by individually processing the power from every battery \cite{Smith2019}.  The disadvantage of FPP is that 100\% of the energy from the battery is processed by an individually dedicated power converter.  FPP requires converters with the highest power rating, and therefore the highest cost of the three architectures we compared.  In applications with FPP the system efficiency directly depends on the power converter efficiency. The trade-off between converter cost and system efficiency is further complicated by the increased cost of thermal management. Therefore, the use of lower cost and lower efficiency converters may not result in a lower overall system cost. To overcome these limitations, we present a power conversion architecture together with a design framework that uses partial power processing to increase system efficiency, reduce converter ratings and hence capital cost, and lessen power processing and hence cost of thermal management.

Partial power processing methods have been applied to batteries including hierarchical active balancing architectures \cite{Zhang2017a}, interleaving the batteries together with the power electronics \cite{Hua2016}, and active balancing architecture using virtual bus \cite{Evzelman2016}. Partial power processing has been shown to increase the energy production and reliability of photovoltaic systems \cite{Shenoy2013a}. Conventional partial power processing (C-PPP), however, to the authors' knowledge has not been used to provide energy buffering in EV charging with second-use batteries. The authors hypothesize that the reason for this is that {\em conventional} partial power processing does not perform well for batteries with a large degree of heterogeneity. Partial power processing has been used in series \cite{Olalla2015} and parallel \cite{Zhou2015} photovoltaic arrays.  Typical C-PPP architectures for new batteries include adjacent battery-to-battery \cite{Hua2016, Kim2014a, Stauth2013, Ye2015}, ac-bus-coupling \cite{Lim2014a, Li2013b, With2015a}, output dc-bus-coupling \cite{Einhorn2011, Mestrallet2014},  and virtual dc-bus-coupling \cite{Evzelman2016}. Several hierarchical architectures for BESS with new batteries were proposed in \cite{Dong2015b, Chen2015b, Zhang2017a}; the performance objectives for these articles were the increase in the speed of balancing rather than solving heterogeneity in energy capacity.

In this paper we develop tools for design of a new Lite\nobreakdash-Sparse Hierarchical Partial Power Processing (LS-HiPPP) architecture, shown in Fig.~\ref{fig:Topo_Comp}(c), that uses sparse power processing, where fewer power converters than batteries are used for the bulk power processing. The reduction in number of converters together with built-in economies of scale by modularizing converters into a small number of discrete power ratings results in lower system costs.  The design framework uses stochastic models for EV charging, time-varying models for grid capability, and statistical models for battery supply heterogeneity.  The design framework results in statistically-optimal energy storage systems with a reduced levelized cost of charging \cite {Borlaug2020}. The methods that are delineated in our paper can be used useful for other energy storage applications, including hybridized power and energy systems that integrate both batteries and supercapacitors \cite{Elmorshedy2021}.

This design paradigm for second-use battery energy storage systems aims to optimize: (i)~production choices like specific converter types (e.g. power ratings) so that they are optimal for the local battery supply statistics; and (ii)~placements of batteries and converters for each individual 2-BESS. For example, consider a two-level power conversion hierarchy shown in Fig.\,\ref{fig:Topo_Comp}, wherein the first level of power conversion is sparse-heavy and second level of power conversion is lite\nobreakdash-dense.
The first level of power conversion handles the average statistical mismatch among batteries, and between the batteries and the output trajectories. This first level topology is designed using multiple objectives and constraints, which renders as a mixed-integer optimization together with the embedded linear programming for energy flow optimization within this combined battery and power converter network.
This step allows the optimal selection of power converter ratings for the battery storage manufacturer in the production of 2-BESS.
This step allows the optimal selection of power converter ratings for the battery storage manufacturer to produce.
The second level of power conversion consists of a set of low power converters to handle the statistical variation among the energy capacity of batteries (e.g. supply deviation), and between the batteries and the output trajectories.  The overall power flow of the entire hierarchy is optimized based on the statistical deviations, from which the second-level converter ratings can be derived and similarly used by the manufacturer to order power converters.

During 2\nobreakdash-BESS production, the set of available converters is pre-determined. The 2-BESS operation occurs when all power converters and batteries have been purchased. During operation, the energy flow of the battery and power converter network within the energy storage system is optimized using linear programming at every time step while ensuring battery and power converter constraints related to power output and battery state of charge are satisfied.  


This article verifies the hypothesis of better performance of a new partial power processing architecture---Lite-Sparse Hierarchical Partial Power Processing (LS-HiPPP), in comparison to C-PPP from literature, in an EV charging application.  The specific list of contributions are as follows:
\begin{enumerate}
\item Quantitative comparisons of LS-HiPPP, FPP, and C-PPP in an EV charging scenario.
\item Formulation of an EV charging test scenario for a fair quantitative evaluation of power processing architectures in a 2-BESS.
\item Optimization of LS-HiPPP for EV charging scenarios.
\item Definition of new statistical metrics
to compare 2-BESS techno-economic performance: {\em derating} and {\em captured value}.
\item Demonstration of improvement for 2-BESS using LS\nobreakdash-HiPPP in captured value, derating, quality of service, energy utilization, and resilience to usage uncertainty.
\end{enumerate}

The paper is organized as follows.
Section\,\ref{sec:Intro} is an overview and literature review of the opportunities, challenges, and previous research in 2\nobreakdash-BESS, partial power processing and EV charging. Section\,\ref{sec:Method} provides the methodology to design and compare LS\nobreakdash-HiPPP to other power processing architectures:
Section\,\ref{sec:2bess_system} delineates the optimization of LS\nobreakdash-HiPPP;
Section\,\ref{sec:charing_scenario} formulates the EV charging scenario that is used for comparison; 
Section\,\ref{sec:ev_charging_model} models the arrival process and energy demand of EVs;
Section\,\ref{sec:3_actor_model_reduction} shows a three-actor model for a singular charging plaza;
Section\,\ref{sec:metric} describes statistical metrics to compare the performance of 2-BESS.
Section\,\ref{sec:results} shows and discusses the results of Monte\nobreakdash-Carlo simulations and quantitatively compares the performance of LS\nobreakdash-HiPPP, C\nobreakdash-PPP, and FPP. Section\,\ref{sec:conclusion} summarizes the key results and suggests extensions and future work.
\section{Methodology} \label{sec:Method}

A new method is presented in this section for using a reduced model for an EV charging plaza to perform unbiased comparisons among second-use battery energy storage systems with different power processing architectures.
Also, a new method for power processing in 2-BESS is discussed together with methods for design and optimization. This new LS-HiPPP method is compared with conventional power processing architectures from literature using scenarios from the reduced EV charging model. The types of statistical distributions used in the scenarios including EV arrival rate \cite{Bayram2021,Lahariya2020}, EV charging energy demand \cite{Cao2012,Arias2016}, and second-use battery heterogeneity in the capacity \cite{Debnath2014,Keeli2012} are identical to those used in the literature.

\subsection{Design of Second-Use Battery Energy Storage Systems} \label{sec:2bess_system}
In literature, the degradation of energy capacity of an ensemble of batteries follows a Gaussian cross-sectional distribution at a particular time slice \cite{Richardson2017, Li2020}.  This is consistent with the energy capacities $E_{b2u}$ of an ensemble of EV batteries destined for second-use following a Gaussian distribution at the time of their retirement \cite{Debnath2014, Keeli2012}. We denote the depth of discharge (DoD) of second-use batteries by a constant $D_{b2u}$. The individual battery energy $E^{b}$ follows
\begin{align}
    E^{b} = E_{b2u}\,D_{b2u}.
\end{align}

\begin{figure}[ht]
    \centering
    \includegraphics[width=6cm]{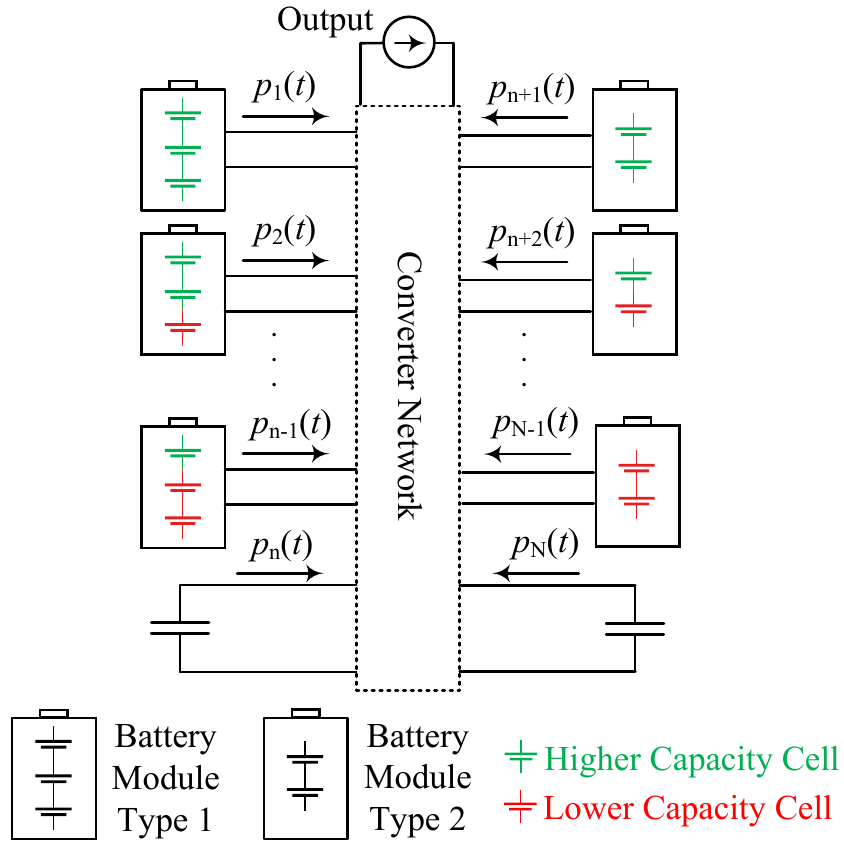}
    \caption{\label{fig:bess_network} Battery storage network design can be formulated as a network energy flow optimization problem.}
\end{figure}

\begin{figure*}
    \centering
    \includegraphics[width=6in]{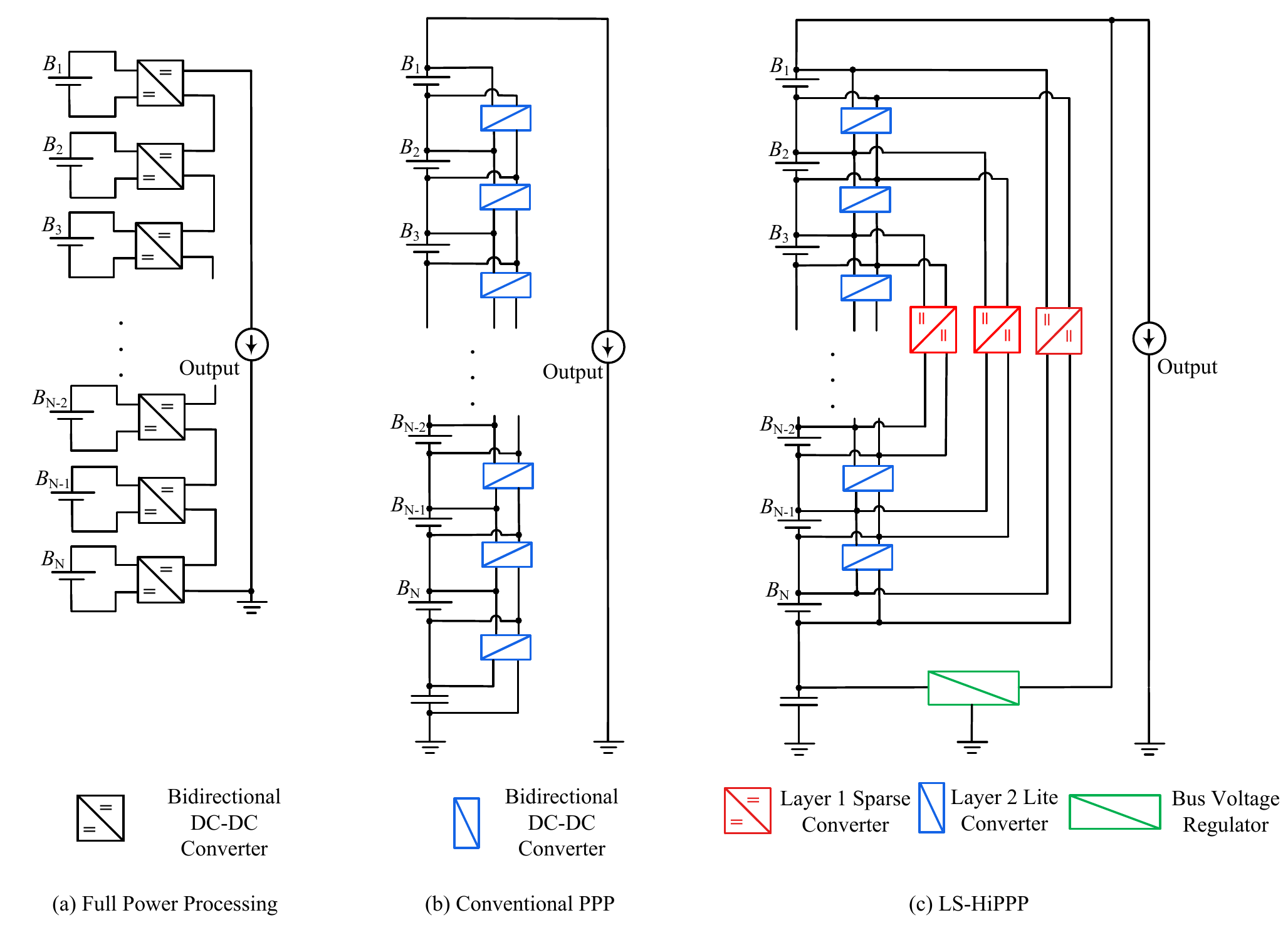}
    \caption{(a) Full Power Processing (FPP); (b) Conventional Partial Power Processing (C-PPP); (c) Lite-Sparse Hierarchical Partial Power Processing (LS-HiPPP). The batteries are interconnected in series together with a bus voltage regulator.}
    \label{fig:Topo_Comp}
\end{figure*}

A battery energy storage system is composed of batteries and power converters that are interconnected as a network with an input-output power port, as illustrated in Fig.~\ref{fig:bess_network}.
We compare our (i) LS-HiPPP architecture to two current state of the art approaches: (ii) FPP and (iii) C-PPP, as illustrated in Fig.~\ref{fig:Topo_Comp}.
The specific configuration for LS-HiPPP for our comparison is a series interconnection of batteries with bidirectional power converters shaping the energy flow as illustrated in Fig.~\ref{fig:Topo_Comp}(c).  Output voltage regulation for LS-HiPPP is performed by a bidirectional power converter; in general, this power converter can output both negative and positive voltages to accommodate different required output voltages and battery voltage heterogeneity.  It is worth noting that accommodating battery voltage heterogeneity is homologous to accommodating different output voltages.



\subsubsection{Battery Storage Network Design} \label{subsec:BSND}
Battery storage network design can be considered as a topology design problem on a directed graph $\mathcal{G}  = (\mathcal{V}, \mathcal{A})$. $\mathcal{V}$ is a \emph{node} set whose elements are the battery modules. $\mathcal{A}$ is the set of ordered pairs of battery modules. The elements of $\mathcal{A}$ are \emph{edges} of $\mathcal{G}$, which represents the electrical interconnection among battery modules. 

The states of each battery module include energy capacity $E^b_i$ and voltage $V^b_i$. To show the statistical benefits of the proposed architecture, the $E^{b}_i$ are modeled as random variables. Their probabilistic distributions $p(E^b_i)$ can be extracted from heterogeneous second-life battery supply statistics. We introduce multiple statistical indicators $\varphi: \mathbb{X} \rightarrow \mathbb{R}$, where $\mathbb{X}$ is the space of random variables, so that we can easily describe the behavior of battery modules and overall 2-BESS.  The statistical indicators $\varphi$ that we consider in this paper include the mean, standard deviation, and interdecile range (IDR) \cite{MichelJambu1991}. 
The energy processing design will be discussed in the following section.

\subsubsection{Energy Processing Design} \label{subsec:EPD} 
Energy processing design is a network topology design and flow calculation problem on a directed graph $\mathcal{G}^s  = (\mathcal{V}^s, \mathcal{A}^s)$. $\mathcal{G}$ in Section\,\ref{subsec:BSND} is a subgraph of $\mathcal{G}^s$ \cite{Bondy1976}. 
The vertices $\mathcal{V}^s \setminus \mathcal{V}$ represents the \emph{node} set whose elements are power converters, input ports, and output ports. 
The edges $\mathcal{A}^s \setminus \mathcal{A}$ is a set of ordered pairs between batteries and power converters, batteries and input/output ports, and power converters and input/output ports. $\mathcal{A}^s$ consists of \emph{edges} of $\mathcal{G}^s$, which represent circuit interconnections. $\mathcal{A}^s$ also contains the \emph{weight} of each edge, which represents the energy flow on the circuit interconnection. 
The power converter node is characterized by its power converter ratings $p_i$ and voltage ratings $v_i$. The nodes which represent the input/output ports contain multiple properties, e.g. power, energy, and voltage.

$\mathcal{G}^s$ is a temporal graph \cite{Bondy1976} in either continuous time, discrete time, or event-driven sampled space.
The time-varying nodal state can cause time-varying weights in $\mathcal{A}^s$. For example, if the input and output node represent the available grid constraint and vehicle charging demand, respectively, the time-varying input/output can cause the energy flow on each converter and batteries to evolve with time. 
Electrical contactors and relays can enable the dynamic connectivity structure \cite{Bondy1976} on $\mathcal{G}^s$. $\mathcal{A}^s$ can be temporarily enlarged and shrunk. $\mathcal{V}^s$ can vary as batteries are placed in and out of service.

In this paper, the node set in the battery storage network $\mathcal{V}$, is randomly sampled from the heterogeneous second-life battery supply statistics, which follow a Gaussian distribution. Therefore the statistical indicators $\varphi$ include the mean and standard deviation. The edge set in battery storage network $\mathcal{A}$ forms a series-string topology.

We focus on a new LS-HiPPP architecture for 2-BESS, whose node and edge sets $\mathcal{V}^s$ and $\mathcal{A}^s$ are shown in Fig. \ref{fig:Topo_Comp}(c). The LS-HiPPP structure clearly exhibits a two-layer energy processing architecture. Layer 1 and 2 can be represented by graph $\mathcal{G}_1$ and $\mathcal{G}_2$ which are subgraphs of $\mathcal{G}^s$.
Layer 1 is considered \emph{sparse} because the number of converter nodes in $\mathcal{G}_1$ (i.e. the order of $\mathcal{G}_1$) is much smaller than the number of battery nodes in $\mathcal{G}$ (i.e. the order of $\mathcal{G}$). 
Layer 2 is considered \emph{dense} because the order of  $\mathcal{G}_2$ is equal to the order of $\mathcal{G}$ minus one.
In Fig.\,\ref{fig:Topo_Comp}(c), Layer 2 is realized by placing a power converter between each battery and its adjacent battery in the string.

The energy design can be divided into two phases. In the first phase, given the battery supply statistics, we need to select the sets of power converters that will be used in the production of 2-BESS units. 
The second phase occurs when all power converters have been purchased and the energy capacities of all batteries have been known. The allocation of and power flow through power converters are decided in second phase. 

\subsubsection{2-BESS Modeling}
The following 2-BESS model makes the 2-BESS optimization tractable because linear programming can be used.  This is important because repeated invocations of linear programming are needed for the 2-BESS mixed-integer optimization of interconnection, and for energy flow optimizations in the Monte Carlo evaluations of performance metrics over ensemble statistics.  This reduced computational complexity allows us to compare different 2-BESS architectures with battery heterogeneity through optimal tradeoff curves.

There are several key approximations that enable the formulation of the energy flow optimization as a linear program: 
\begin{itemize}
    \item Zero power converter losses.
    \item Zero battery losses.
    \item Quasistatic available grid power $P_{ag}[n]$ for the $n^{\text{th}}$ charging cycle.
    \item Quasistatic EV charging energy demand.
    \item Singular depletion model \footnote{{\em Singular depletion model} for 2-BESS means that the 2-BESS is considered to be depleted when one battery is depleted. } for 2-BESS.
    \item Each battery voltage can be different, but time-invariant over discharge.
\end{itemize}

The optimality gap from power converter and battery losses is expected to be small.  Partial power processing architectures that include LS-HiPPP and C-PPP are designed to process only the mismatch power. Typically less than 20\% of the system output power is processed. Even if lower efficiency and therefore lower cost converters are used, the overall system losses can be small, as shown in Fig.\,\ref{fig:efficiency_rating}.  For example, if power converters with an efficiency of 85\% are used in a system where only 15\% of the power is processed, then the losses are only 2.25\% of the system output power. Battery losses are expected to be even smaller than the converter losses \cite{Mongird2019}.

The 2-BESS energy is considered depleted when a chosen number of battery modules reaches it depth of discharge limit. In a series configuration, when a battery is depleted, it is bypassed, causing the battery stack voltage to drop.  This requires the bus voltage regulator to process more power to accommodate this voltage drop, hence causing the system efficiency to decrease and also requiring this power converter to have a higher rating. Second-use batteries have the longest lifetime when operated in the state of charge where the battery voltage is nearly constant \cite{Saxena2016}. To reduce complexity, we choose the threshold for depletion to be when one battery is depleted, i.e. {\em singular depletion model} for 2-BESS.

\subsubsection{Power Conversion Design}
Power conversion design is a network topology design and flow optimization problem. The design variables in Layer 1 include power converter ratings, denoted by $p^{(1)}_1, \ldots, p^{(1)}_M$ and  the number of converters, denoted by $M$.  The design variables in Layer 1 also include the interconnections between batteries and power converters, i.e. set $\mathcal{A}^s \setminus \mathcal{A}$.
The design variables in Layer 2 include power converter ratings of converters, denoted by $p^{(2)}_1, \ldots, p^{(2)}_{N-1}$. The number of power converters in Layer 2 is pre-selected to be one less than the number of batteries. The interconnections between batteries and power converters are predetermined as the adjacent module-to-module structure shown in Fig.\,\ref{fig:Topo_Comp}(c).

In the battery discharging scenario, each battery module is considered as the energy source with capacity $E^b_j$. The load is modeled as the energy sink $E_{\text{out}}$.
The optimization objective is to maximize the energy utilization of the 2-BESS, defined as the ratio of energy delivered to the load $E_{\text{out}}$ and the summation of the available energy sources
\begin{align} \label{eqn:bat_energy_utilization}
    \mathcal{U_E} \triangleq \frac{E_{\text{out}}}{\sum\limits E^b_j}.
\end{align} 
The power ratings can be selected by averaging the energy over the discharging period. This required discharge time is determined by the EV charging application.

Our design utilizes two well-accepted techno-economic approximations: (i) The cost of a power converter is nearly proportional to its power rating. Smaller sum of  power converter ratings results in a lower cost \cite{Mongird2019}; and
(ii) Fewer types and larger numbers of power converters are favorable for economies of scale.
The power ratings of Layer 2 converters are chosen to be identical; the Layer 1 converters are chosen to be identical, but in general different from Layer 2.
The detailed design procedure in Layer 1 and 2 will be discussed in the next section.
 


\subsubsection{Power Processing Design Using Energy Flow Optimization}\label{sec:pp_optimization}

\paragraph{Layer 1 Power Processing Design} \label{sec:layer1_ppp}
The power processing design of Layer 1 is a mixed-integer stochastic optimization problem because the battery storage network is stochastic and the design variable $\mathcal{A}^s \setminus \mathcal{A}$ is combinatorial.
We convert the stochastic optimization problem to a  deterministic optimization problem by a Distribution Flattening method \cite{Cui2021a}. 
Given an arbitrary probabilistic distribution of the battery capacity $p(E^b_i)$ and number of batteries $N$, the Distribution Flattening method can generate the expected set $\mathcal{B} = \{B_1, B_2, \cdots, B_N\}$, the $N$-element set of batteries that represent the expected performance.
In the series-string battery storage network discussed in this paper, $B_1 \ldots B_N$ are placed from lowest to highest expected energy $\bar{E}^b_1 \cdots \bar{E}^b_N$.

Given a small number of batteries and a sparse arrangement of power converters, we exhaustively search the set of feasible interconnection, under the constraint of converter voltage ratings. For every feasible interconnection, the optimal energy flow is found using linear programming
\begin{align}
\underset{e^{(1)}_i,\,e^b_{j},\,e^{\text{out}}_{j}}{\mathrm{max}}& \,\,\,  \sum\limits_{1\le j \le N} e_j^{\text{out}} \label{eqn:opt_layer1_obj} \\
\text{subject to} & \,\,\,\ e^{\text{b}}_j \le \bar{E}^b_j \label{eqn:opt_layer1_contraint1},\\
& e^{\text{b}}_j = \sum\limits_{i \in K^{(1)}_j} e^{(1)}_i + e_j^{\text{out}}, \label{eqn:opt_layer1_contraint2} \\
& e_j^{\text{out}} = Q_{\text{string}}V^{b}_j, \quad j = 1,\,2,\,\ldots,\,N,\label{eqn:opt_layer1_contraint3}
\end{align}
where $e_j^{\text{out}}$ is the energy delivered from the $j^{\text{th}}$ battery to the output, $e_j^{\text{b}}$ represents the output energy of the $j^{\text{th}}$ battery, $K^{(1)}_j$ is the index set of the Layer 1 converters whose inputs are connected to $j^{\text{th}}$ battery.

The design space consists of the energy processed by $M$ Layer 1 converters, denoted by $e^{(1)}_1 \cdots e^{(1)}_M$. The optimization objective (\ref{eqn:opt_layer1_obj}) is to maximize the total energy delivered to the output. Constraint (\ref{eqn:opt_layer1_contraint1}) limits the available battery energy capacity. The constraint indicates that the entire 2-BESS is considered to be depleted if any battery module is depleted. Constraint (\ref{eqn:opt_layer1_contraint2}) represents the energy conservation law for each battery. 
Constraint (\ref{eqn:opt_layer1_contraint3}) indicates the direct energy transfer from the $j^{\text{th}}$ battery to the output linearly depends on battery voltage $V^{b}_j$ because the charge through the whole battery string is the same $Q_{\text{string}}$.

Finally, the maximum energy output $ E_{\text{out}}^{*}$, the corresponding optimal circuit topology $K^{(1)^{*}}_j$, and optimal energy flow $e^{(1)^*}_1 \cdots e^{(1)^*}_M$ are selected by maximizing battery energy utilizations among all feasible typologies.

The optimal set of processed power can be obtained by averaging the optimal processed energy by the required discharging period $T$ defined in (\ref{eqn:agg_conv_rating})
\begin{align}
    p^{(1)^*}_i = \frac{e^{(1)^*}_i}{T}, \quad i = 1,\,2,\,\cdots,\,M.
\end{align}

The Layer 1 converter rating is selected to be the maximum of the optimal set of the processed power.

\paragraph{Layer 2 Power Processing Design}
\label{sec:layer2_ppp}
Compared to the goal of maximizing expected performance in Layer 1, Layer 2 aims at processing the mismatch energy from the statistical deviation of the batteries from the expected set. 
Layer 2 power processing design is a stochastic optimization problem. Monte Carlo methods are exploited to optimize over statistical deviations.

The design of Layer 2 follows the design of Layer 1. The optimal interconnection, maximum energy processed, and power ratings of the Layer 1 converters are treated as the constraints in the design of Layer 2. The optimization goal is to determine the power rating $p^{(2)}_{\text{max}}$.

The optimization starts by setting the \emph{hierarchical parameter} $\lambda_H$, defined as the ratio of the aggregate power rating between the two layers, 
\begin{align}
    \lambda_H \triangleq \frac{\sum\limits_{i = 1}^{N-1} p^{(2)}_{\text{max}}}{\sum\limits_{i = 1}^{M} p^{(1)^{*}}_{i}}.
\end{align}
Smaller $\lambda_H$ indicates a stronger hierarchical power processing architecture. By setting $\lambda_H = 0$, LS-HiPPP is diminished to sparse partial power processing. By increasing $\lambda_H$, LS-HiPPP converges to C-PPP. As $\lambda_H \rightarrow \infty $, power processing is dominated by Layer 2, which is identically interconnected like C-PPP.
Each $\lambda_H$ determines the power rating of the Layer 2 converters. The battery energy utilization can consequently be calculated by the following linear programming problem
\begin{align}
\underset{e^{(1)}_i,\,e^{(2)}_{k},\,e_j^b,\,e^{\text{out}}_j }{\mathrm{max}}& \,\,\, \sum\limits_{1\le j \le N} e_j^{\text{out}} \label{eqn:opt_layer2_obj} \\
\text{subject to} & \,\,\,\ e^{\text{b}}_j \le (\bar{E}^b_j + \delta E^b_j) \label{eqn:opt_layer2_contraint1},\\
& e^{\text{b}}_j = \sum\limits_{i \in K^{(1)^{*}}_j}  e^{(1)}_i + \sum\limits_{k \in K^{(2)}_j} e^{(2)}_k + e_j^{\text{out}}, \label{eqn:opt_layer2_contraint2}\\
& e_j^{\text{out}} = Q_{\text{string}}V_j^{b}, \quad j = 1,\,2,\,\ldots,\,N,\label{eqn:opt_layer2_contraint3}\\
& e^{(2)}_k \le \frac{\lambda_H \sum\limits_{i = 1}^{M} e^{(1)^{*}}_{i}}{(N-1)}, \quad k = 1,\,2,\,\ldots,\,N-1,\label{eqn:opt_layer2_contraint4}\\
& e^{(1)}_{i} \le e^{(1)^{*}}_{i}, \quad i = 1,\,2,\,\ldots,\,M \label{eqn:opt_layer2_contraint5},
\end{align}
where $K^{(2)}_j$ is the index set of the Layer 2 converters whose inputs are connected to the $j^{\text{th}}$ battery. 

The design space consists of the energy processed by $N-1$ Layer 2 converters, represented by $e^{(2)}_1 \ldots e^{(2)}_{N-1}$.  The optimization objective in (\ref{eqn:opt_layer2_obj}) is to maximize the total energy transferred to the output, given the ratings and interconnection of Layer 1 and Layer 2 converters.
In Constraint (\ref{eqn:opt_layer2_contraint1}), the extra $\delta E^b_j$ represents the energy deviation of the $j^{\text{th}}$ battery to the expected value $\bar{E}^b_j$. Constraint (\ref{eqn:opt_layer2_contraint2}) represents the energy conservation for each battery. Constraint (\ref{eqn:opt_layer2_contraint3}) expresses the direct energy transfer from the $j^{\text{th}}$ battery to the output. Constraint (\ref{eqn:opt_layer2_contraint4}) indicates the power ratings for all Layer 2 converters identically equal to $p_{\text{max}}^{(2)}$. Constraint (\ref{eqn:opt_layer2_contraint5}) enforces that the Layer 1 converter ratings are kept fixed in the Layer 2 power processing design. 

For each Monte Carlo sample from the battery supply distribution, an optimal battery energy utilization is obtained. The Monte Carlo method results in a distribution of optimal battery energy utilizations. The mean and standard deviation as well as other statistical indicators can be calculated from this distribution and designated to the secondary converter rating $p_{\text{max}}^{(2)}$. 

By changing the hierarchical parameter $\lambda_H$, the tradeoff curve between the Layer 2 converter rating and battery energy utilization metric is obtained. 


\subsubsection{Deployment of 2-BESS Units}
In production, all the power converters are procured in mass quantities, which are determined by the 2-BESS product demand. For each 2-BESS, the batteries are sorted from the lowest to highest in power capability and then connected in series. The power converters are connected according to section~\ref{sec:pp_optimization}.

The optimal power flows are recalculated using linear programming during operation given the actual battery energy capacities in the 2-BESS unit and the output load. As we show in Section \ref{sec:results}, the design method for LS-HiPPP results in a better cost-performance tradeoff as it relates to battery energy utilization and normalized aggregate converter rating in comparison to competing approaches.  2-BESS with an LS-HiPPP architecture performs well with a sparse set of Layer 1 converters at moderate power ratings and a dense set of Layer 2 converters at lower power ratings.
\subsection{Formulation of EV Charging Plaza Scenario} \label{sec:charing_scenario}
The control and optimization of EV charging microgrids with energy storage is complex and an active research topic \cite{Yang2019a, Faisal2018}.  
Also, power processing for battery energy storage systems has been studied \cite{Zhang2017a}. However, a comparison of the performance of full power and partial power processing architectures with second-use battery energy storage systems to the authors' knowledge has not been previously investigated for EV charging applications.  By simplifying the model for EV charging demand and grid supply, we can unravel some of the complexity by deconflating the effects of power processing architecture from the allocation and scheduling of multiple storage systems and multiple EV chargers \cite{Wang2019a, Bryden2019, Wang2020c}.

In this paper, several abstractions are needed to simplify the modeling of the interactions between the grid, EV charging, and second-use battery energy storage.  These simplifications are manifested in the modeling of grid constraints, charging demand and operation, and battery energy storage system design and operation.  These models allow us to compare different 2\nobreakdash-BESS architectures by enabling us to untangle the effects on performance of multiple stochastic and statistical constraints.

The charging plaza as as illustrated in Fig.\,\ref{fig:new-charging-plaza} consists of three coordinated controllers: (i) Grid Interface Controller (GIC); (ii) Charging Management System (CMS); and (iii) BESS Management System (BeMS).  The GIC determines the ultimate power constraints for the dc microgrid.  The grid operator communicates with the GIC to set the grid capability at the connection point, which is determined by physical and electrical constraints, and other loads.  The CMS controls power partitioning among EV chargers, charger availability, and scheduling.  The CMS can also control the variable pricing for EV charging, which is indirectly related to demand response pricing controls from the grid.  The BeMS interfaces with individual 2-BESS units.  The BeMS treats each 2-BESS as an energy storage monolith with a well-defined data and information interface; this interface exposes aggregate parameters and external states, such as external state of charge, as an abstraction.  The BeMS controls power partioning among the 2-BESS units, external depth of discharge, recharging, scheduling, and use leveling, among others. The GIC, CMS, and BeMS coordinate their optimization and control efforts for the best performance of the EV charging plaza.

In practice, an EV charging plaza with 350\nobreakdash-450\,V chargers will employ 50\,V modules extracted from second-use EV battery packs in a series connection.  In the following sections, these modules will be interchangeably be referred to as batteries.
\subsection{EV Charging Model} \label{sec:ev_charging_model}
In literature, the EV arrival rate $\lambda$ to a charging lot follows a Poisson process \cite{Bayram2021, Lahariya2020}.  In this paper, we use a standby time $T_d$ that follows an exponential distribution with rate $\lambda$, which is the interarrival time of this Poisson process \cite{chung2012elementary}.  The expected standby time
\begin{align}
   \bar{T}_d = \frac{1}{\lambda}.
\end{align}

Also in literature, the initial battery state of charge (SoC) of an EV before its charging $\text{SoC}_{\text{initial}}$ is a random variable that follows a Gaussian distribution \cite{Cao2012, Arias2016}.  In this paper, we denote the energy capacity and end-of-charging $\text{SoC}$ of an EV by $C_{ev}$ and $\text{SoC}_{\text{end}}$, respectively. Therefore, $E_{ev}$, the EV charging energy demand satisfies
\begin{align}
    E_{ev} = C_{ev}(\text{SoC}_{\text{end}} - \text{SoC}_{\text{initial}}).
\end{align}
Because $E_{ev}$ is affinely transformed from $\text{SoC}_{\text{initial}}$, $E_{ev}$ follows a Gaussian distribution \cite{chung2012elementary}.
\subsection{Three-Actor Model Reduction}
\label{sec:3_actor_model_reduction}

In this paper, we examine a simplified case with three actors as a singular charging plaza: (i) single source of power generation, i.e. allocated grid power; (ii) single EV charger; and (iii) single 2-BESS.  We abstract the grid interface as a time-varying available grid power $P_{ag}(t)$, which represents the portion of the grid power that is allocated to the singular subset of one charger and one 2-BESS. We also apply a quasistatic approximation to some of the trajectories, meaning that the variable is constant over a particular charge cycle time interval $t_{ev}[n]$ shown in Fig.~\ref{fig:trajectory_carton}.

This model reduction allows us to perform comparisons of 2-BESS architectures with multiple deterministic time-varying and stochastic constraints over different parameter ensemble statistics. These stochastic constraints model the expected behavior of different aspects of the EV chargers, and implicitly the deviation from the expected behavior. The heterogeneity in the energy capacity of the battery supply is modeled statistically and parameterized by expected values and deviations from the expected values.  

The available grid power to the 2-BESS and EV charger embeds the actions of the GIC and BeMS to the demand response controls from the grid.  The available grid power is modeled as a constraint that is time-varying and deterministic.
The EV charging demand trajectory embeds the stochastic EV arrivals as an interarrival time that manifests as a standby time $T_d$ that has an exponential distribution. 
{\em EV charging energy demand} $E_{ev}[n]$ is quasistatic over the charging time interval $n$ and is represented by a Gaussian distribution, which is parameterized by an expected value and a standard deviation.  The EV charger constraints are parameter specifications, inputs, outputs, and control policy:
\begin{enumerate}[label=(\alph*)]
    \item Parameter Specification: maximum charging power;
    \item Input: available grid power;
    \item Input: available 2-BESS power;
    \item Input: standby time;
    \item Output: {\em EV charging energy demand}, i.e. the energy required by the vehicle per EV charging.
    \item Control policy detailed below.
\end{enumerate}

A statistical model for the second-use battery energy storage system is employed to represent the battery supply heterogeneity. 
We use a Gaussian probability distribution for the battery energy capacity, parameterized by an expected value and a standard deviation that represents the supply heterogeneity. 
The 2-BESS constraints are:
\begin{enumerate}[label=(\alph*)]
    \item Maximum 2-BESS output/discharge power is equal to the maximum charger power;
    \item Maximum 2-BESS output energy  per EV charging cycle is determined by the energy capacity or a specified depth of discharge;
    \item Maximum recharge power is equal to the maximum grid power.
\end{enumerate}

The 2-BESS output energy $E_{\text{out}}$ is the energy delivered by the 2-BESS during a particular time interval, e.g. charging cycle.

The aggregate power converter rating is
\mbox{$\mathcal{P} = \sum p_i$} and the 2-BESS intrinsic energy capacity is \mbox{$ \mathcal{E}^b = \sum E^b_j$}, where $p_i$ is the individual converter rating and $E^b_j$ is the individual battery energy.
The energy-normalized aggregate power converter rating 
\begin{align} \label{eqn:agg_conv_rating}
    \mathcal{R} \triangleq \frac{\mathcal{P}}{\mathcal{E}^b}T,
\end{align}
where $T$ is the time required to deplete the 2-BESS intrinsic energy capacity $\mathcal{E}^b$ of the battery at $\mathcal{P}$.


\subsubsection{Simplified Control Policy for Three Actors}
\begin{figure}[ht]
    \centering
    \includegraphics[width=2.5in]{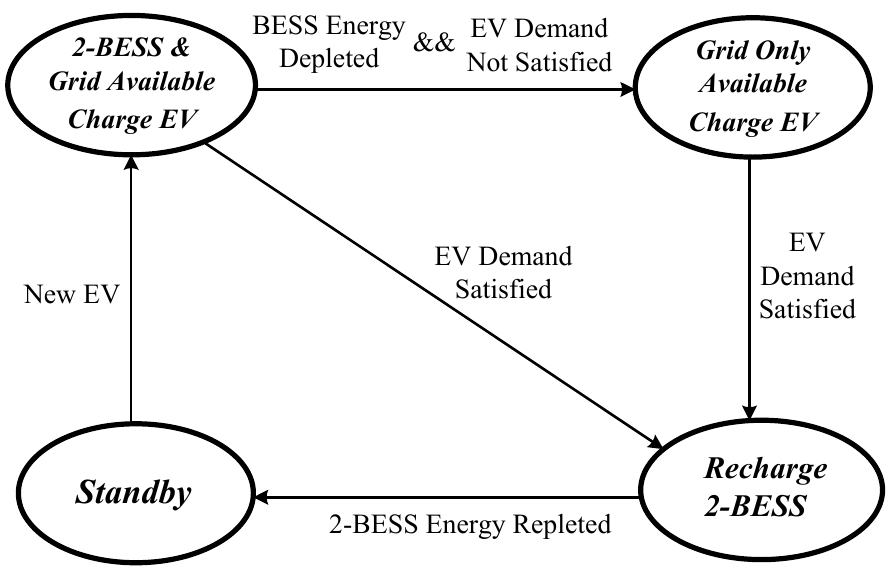}
    \caption{\label{fig:state_diagram} State Diagram of Three-Actor Control Policy.}
\end{figure}

\begin{figure}[ht]
    \centering
    \includegraphics[width=2.5in]{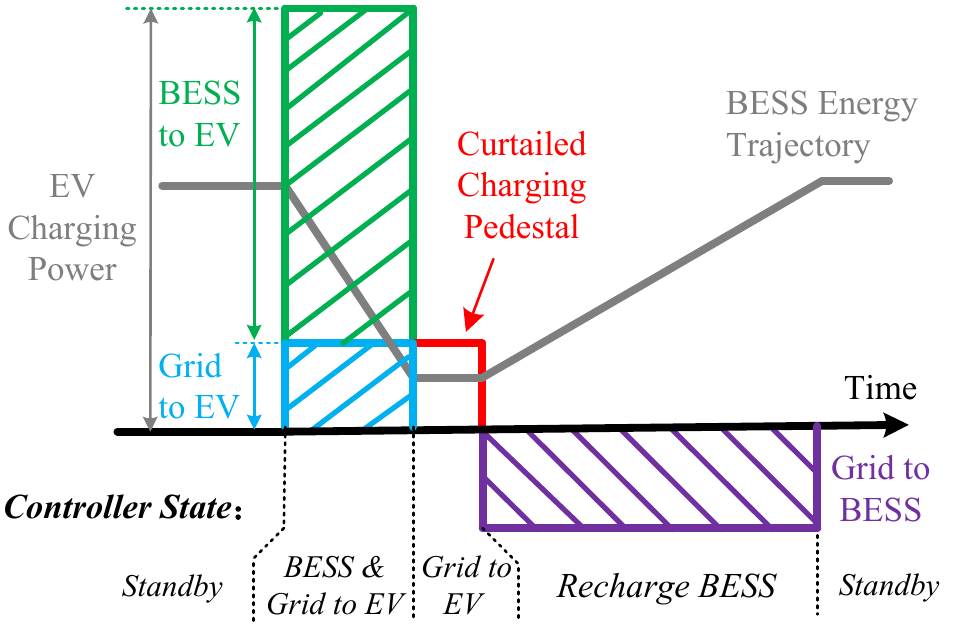}
    \caption{Trajectory of Three-Actor Plaza for an EV charging cycle. The $n^{\text{th}}$ charging cycle has a time interval of $t_{ev}[n]$.} \label{fig:trajectory_carton}
\end{figure}

The control policy for the three actors is illustrated in Fig.~\ref{fig:state_diagram}.  The following properties are enforced:
\begin{itemize}
    \item Both the 2-BESS and available grid power are used to charge an EV unless the 2-BESS energy is depleted.
    \item If the 2-BESS is depleted and then only the grid is used for EV charging at this {\em curtailed charging} level.
    \item The EV charger operates either at its maximum power or the curtailed charging power.  Note that the curtailed charging power is the available grid power for the EV charging interval.
    \item The 2-BESS is recharged completely after every EV charging cycle.
    \item The 2-BESS recharge is the quasistatic available grid power for the EV charging interval.
    \item The available grid power is constant, i.e. quasistatic, during the entire EV charging cycle.
    \item The standby time between EV charging cycles is nonzero.
    \item The charger is infrequently used and there is no queue for EVs. If an EV arrives when the charger is not in the standby mode, the EV will leave the charging plaza.
\end{itemize}
The resulting trajectory of the system is shown in Fig.~\ref{fig:trajectory_carton}.  At the beginning of the EV charging cycle, the 2-BESS energy is replete and the EV can be charged from both the 2-BESS and the grid at the maximum charging power.  If the 2-BESS energy depletes, the curtailed charging results in a pedestal in the charging power, which continues until the EV is charged.  After the EV is charged, the 2-BESS is completely recharged before it is ready for the next EV charging cycle.  After the 2-BESS is replete, there is a finite standby time before the next EV charge cycle.

\subsection{Metrics for Comparisons} \label{sec:metric}
\subsubsection{Battery Energy Utilization}
The battery energy utilization $\mathcal{U_E}$ in (\ref{eqn:bat_energy_utilization}) is a metric for the amount of battery energy internal to the 2-BESS that is available to the output port.  As previously mentioned, the 2-BESS is considered depleted when any battery has reached its depth of discharge limit.

\subsubsection{System Efficiency}
The system efficiency is calculated by assuming all the power converters in the LS-HiPPP, C-PPP, and FPP designs have identically flat efficiencies. This approximation enables a comparison among the 2-BESS power processing approaches.
The system efficiency can be calculated by  
\begin{align}
    \eta_s = 1 - (1 - \eta_c) \mathcal{R}, 
\end{align}
where $\eta_c$ is the efficiency of all power converter and $\mathcal{R}$ is the normalized aggregate converter rating. For partial power processing approaches, as power converter ratings are reduced, less power is processed, hence a higher system efficiency, albeit at lower battery utilization.  FPP has a constant system efficiency because all the output power is necessarily processed.


\subsubsection{Ensemble Performance Over Single-Day Trajectory}
A single-day time series for the three-actor plaza illustrates the dynamic behavior of the EV charger, grid, and 2-BESS. Together, the available grid power (Grid Constraint), 2-BESS power output (2-BESS Power), 2-BESS remaining energy (2-BESS Energy), and EV charging power comprises the {\em trajectory set}. The exemplar used for comparison in this paper is a single-day time series.  The available grid power and EV charging power comprise the {\em external trajectory set} that determines the behavior of the 2-BESS, represented by the 2-BESS power and energy, which comprise the {\em 2-BESS trajectory set}.

The statistical performance of LS-HiPPP and C-PPP over battery energy heterogeneity will be evaluated.  Monte Carlo simulations were performed with individual battery energy capacity as the random variable.  The output random variable is battery energy utilization. This output random variable is drawn from the statistics of the charging interval.  For interval~$n$, the interval statistical distribution is $\mathcal{Q}_n$.

The {\em grid-EV energy gap} is the difference between the EV charging energy demand and the available grid energy.  The available grid energy $E_{ag}[n]$  is the quasistatic available grid power integrated over the germane charging time interval $n$, \mbox{$E_{ag}[n] = P_{ag}[n]\,t_{ev}[n]$}.

\subsubsection{Statistical Dispersion of Performance}
The statistical measures of dispersion of a performance metric are an indicator of the unit-to-unit variability in the production of 2-BESS for a particular battery heterogeneity.
The IDR (Inter-Decile Range) and standard deviation are two widely used metrics.
Lower IDR and standard deviation are desired because they mean higher consistency of performance when 2-BESS is needed.

\subsubsection{Resilience to Usage Uncertainty}
By examining the effect of the stochastic deviation in the normalized grid-EV energy gap on expected battery energy utilization, we can compare the resilience of LS-HiPPP and C-PPP to uncertainties in usage.

\subsubsection{Curtailed Charging from Usage Uncertainty}
When the stochastic deviation from expected usage is high, the 2-BESS is more likely to deplete in energy, resulting in curtailed EV charging.  {\em Curtailed EV charging} occurs when the 2-BESS is depleted and EV charging power is reduced to only what the grid can instantaneously supply. 

\subsubsection{Derating and Confidence of Performance}
The confidence in the unit-to-unit performance of a 2-BESS can be inferred from statistical measures of dispersion.  For example, a $3\sigma$ confidence in the 2-BESS energy capacity means that there is a 99.85\% probability for a normal distribution that the deployed 2-BESS units will have at least this capacity.

A typical approach to achieving the required confidence in a level of performance is to derate the unit. It is worth noting that derating can be applied to impose a confidence level to any performance metric, not just energy capacity.

\subsubsection{Captured Value of the 2-BESS}
Both derating factor $D_f$ and energy utilization $\mathcal{U_E}$ affect the value of the 2-BESS as a means of energy storage.  The captured value can be defined as the product of the derating, energy utilization, and expected intrinsic energy capacity $\bar{E_I}$
\begin{equation}
    C_v \triangleq D_f\,\mathcal{U_E}\,\bar{E_I},
\end{equation}
where 
\begin{align}
    \bar{E_I} = \sum\limits_{j = 1}^{N} \bar{E}^b_{j}.
\end{align}

\subsubsection{Monte-Carlo Simulation Methods}
The tradeoffs for several performance metrics were evaluated using Monte Carlo methods to compare LS-HiPPP to current state-of-the-art approaches for power processing in 2-BESS. 

The numerical analyses were performed on a high-performance computing cluster with 12 Intel(R) Xeon(R) CPU E7-8860 processors operating at 2.27\,GHz, each with 32 GB of memory using the Parallel Computing Toolbox in Matlab. Five different battery heterogeneities from $5\%$ to $25\%$ were investigated; this heterogeneity refers to the energy capacity variation among the population of second-use batteries. A Gaussian distribution for battery energy was used, and the heterogeneity defined as the ratio of the standard deviation to the mean.

Layer 1 power processing was designed through the distribution flattening method and optimization of the power converter interconnection and rating. The interconnection was optimized by exhaustively searching a space of approximately ten thousand points with a computation time of approximately 5 hours.

For Layer 2 power processing, twenty normalized aggregate power converter ratings from $10\%$ to $100\%$ were investigated.  One hundred sets of battery energy capacities were sampled from the Gaussian distribution.
Each battery set contains nine batteries, which are connected in ascending order of energy capacity. For each battery set and interconnected power converters, the $linprog$ function in Matlab was used to solve the energy flow optimization problem to obtain the maximum energy utilization. The computational time was approximately 5 hours.

We simulated the 24-hour operation of the charging station with grid constraints.
Sixty stochastic scenarios of EV demands and grid constraints were investigated with a computational time of approximately 10 hours.  The inter-arrival times of EV charging were modeled as Poisson processes; three sets of data with different expected values (0.5\,h, 1.5\,h, and 2.5\, h) were investigated.  The power demand from EV charging was modeled as a Gaussian process; two sets of data with different expected values (33\,kWh and 50\,kWh), each with ten subsets parameterized by different standard deviations ranging from 5\,kWh to 25\,kWh.  One thousand trajectories are sampled from a combination of stochastic processes consisting of the different expected values for inter-arrival, and expected values and standard deviations for power demand of EV charging.  Each trajectory of charging station operation is scheduled from linear optimization.

The following specifications are enforced:
\begin{itemize}
    \item Maximum EV charging power is 150\,kW, which is equivalent to a Level III charging rate.
    \item Expected 2-BESS power capability is 150\,kW.
    \item Battery energy heterogeneity is a Gaussian distribution with a mean of 37.5\,kWh and a standard deviation of 9.4\,kWh, corresponding to a 25\% heterogeneity.
    \item Nine battery modules are connected in series.
    \item Identical aggregate power converter ratings are used to compare LS-HiPPP to C-PPP and FPP.
    \item Three identical Layer 1 power converters are used for LS-HiPPP.
    \item Eight identical Layer 2 power converters are used for LS-HiPPP.
    \item Individual power converters within a C-PPP 2-BESS have identical power ratings.
    \item Individual power converters with a FPP 2-BESS have identical power ratings.
    \item Identical power converters are used for bus voltage regulation in LS-HiPPP, C-PPP, and FPP.
    \item The bus voltage regulator absorbs all the battery voltage heterogeneity.
\end{itemize}

For a fair comparison between LS-HiPPP and C-PPP, the following are enforced:
\begin{itemize}
    \item Identical sampling of batteries and hence identical individual capabilities and capacities.
    \item Identical battery interconnection.
    \item Identical EV charging energy demand trajectory.
    \item Identical trajectory of available grid power.
\end{itemize}
\section{Results and Discussions} \label{sec:results}
\subsection{Battery Energy Utilization}
\begin{figure}[t]
    \centering
    \includegraphics[width=3 in]{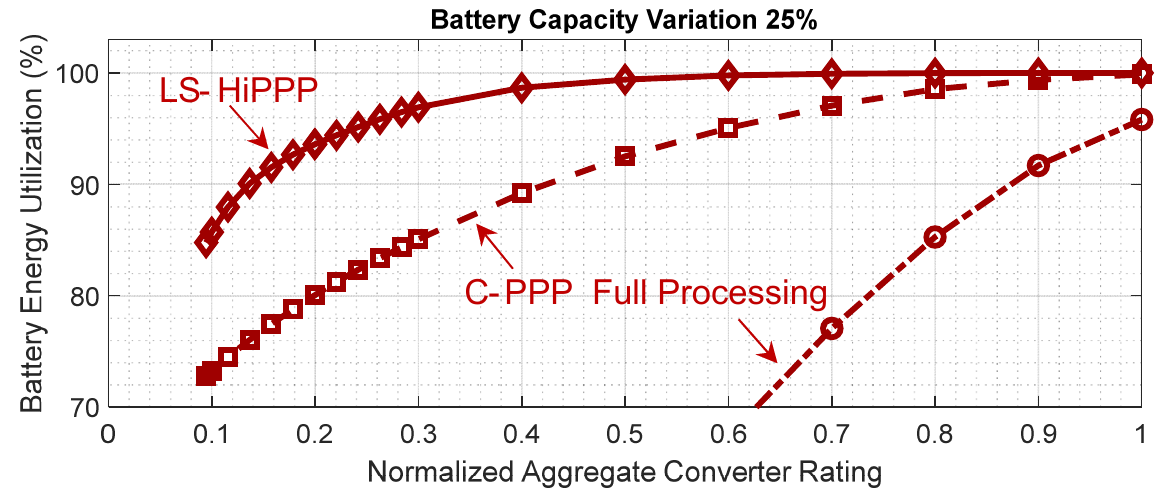}
    \caption{\label{fig:utilization_rating} Comparison of Battery Energy Utilization at 25\% Battery Capacity Deviation for LS-HiPPP, C-PPP, and FPP.}
\end{figure}

The battery energy utilization $\mathcal{U_E}$ is a metric for the battery energy internal to the 2-BESS that is available to the output port. Fig.\,\ref{fig:utilization_rating} shows a comparison of the utilization for LS-HiPPP, C-PPP, and FPP for different aggregate converter ratings at 25\% deviation in battery energy.

Lower aggregate power converter ratings mean lower cost. Fig.~\ref{fig:utilization_rating} shows that LS-HiPPP has significantly better $\mathcal{U_E}$ than both C-PPP and FPP at lower aggregate power converter ratings.  At 20\% aggregate power converter rating, LS-HiPPP has a battery utilization of 94\% versus 78\% for C-PPP; FPP is below the scale.

\subsection{System Efficiency}
\begin{figure}[ht]
    \centering
    \includegraphics[width=3 in]{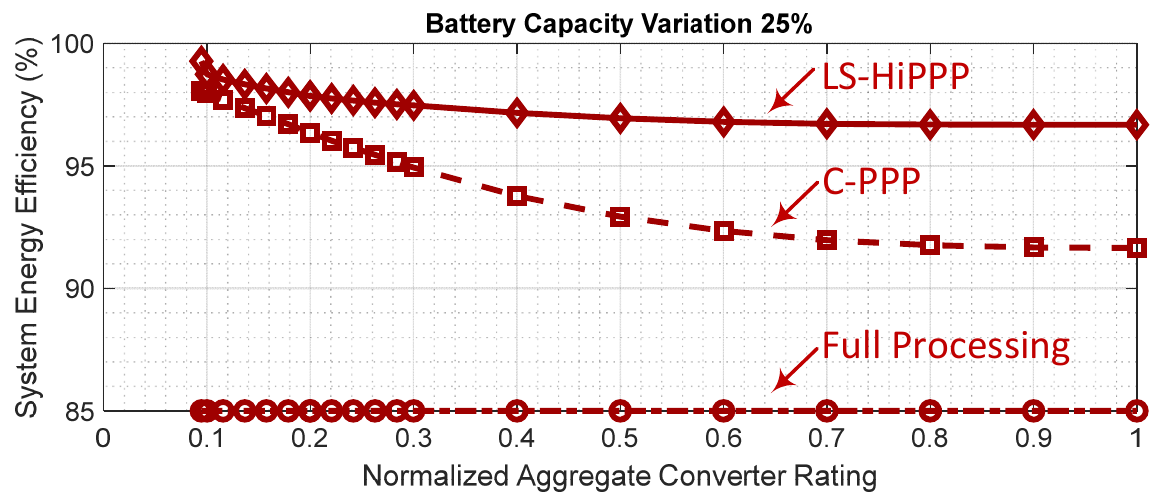}
    \caption{\label{fig:efficiency_rating} System Efficiency at 25\% Battery Capacity Deviation for LS-HiPPP, C-PPP, and FPP for power converters with an 85\% efficiency that is flat.}
\end{figure}

The system efficiency is shown Fig.~\ref{fig:efficiency_rating} for power converters in LS-HiPPP, C-PPP, and FPP designs having identically flat efficiencies. At lower aggregate converter ratings the system efficiency is higher for partial power processing architectures (LS-HiPPP and C-PPP) because less power is processed through the power converter inefficiency.  This higher system efficiency at lower converter ratings has lower battery utilization as shown in Fig. 6.  The system efficiency approaches saturation at high aggregate converter ratings because the battery utilization and hence battery output also approach saturation together with the processed power.  LS-HiPPP has the highest asymptotic system energy efficiency because it processes the least power.

\subsection{Single-Day Time Series for Three Actor Plaza}\label{sec:single-day-3actor}
\begin{figure}[!t]
    \centering
    \subfloat[LS-HiPPP]{\includegraphics[width=3in]{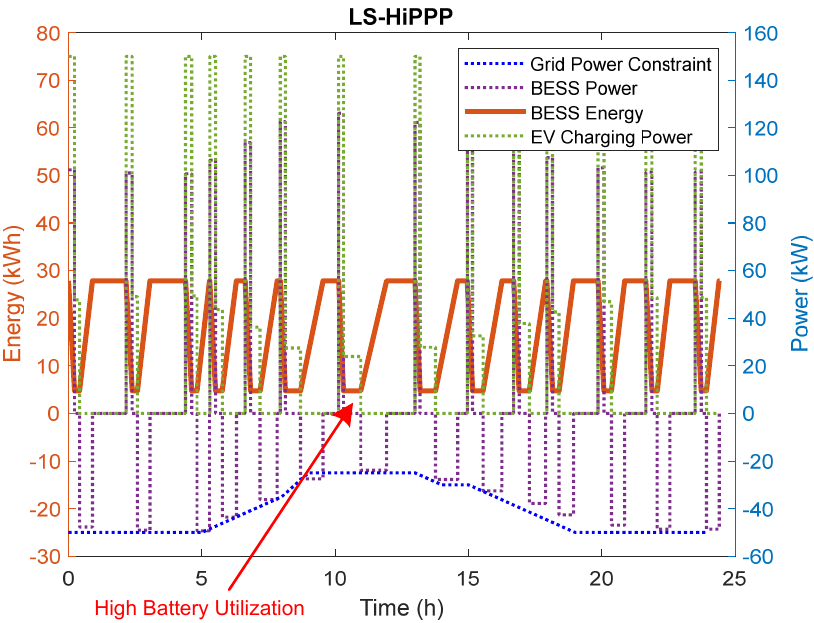} \label{fig:ls-hippp-timeseries}}
    \hfil
    \subfloat[C-PPP]{\includegraphics[width=3.1in]{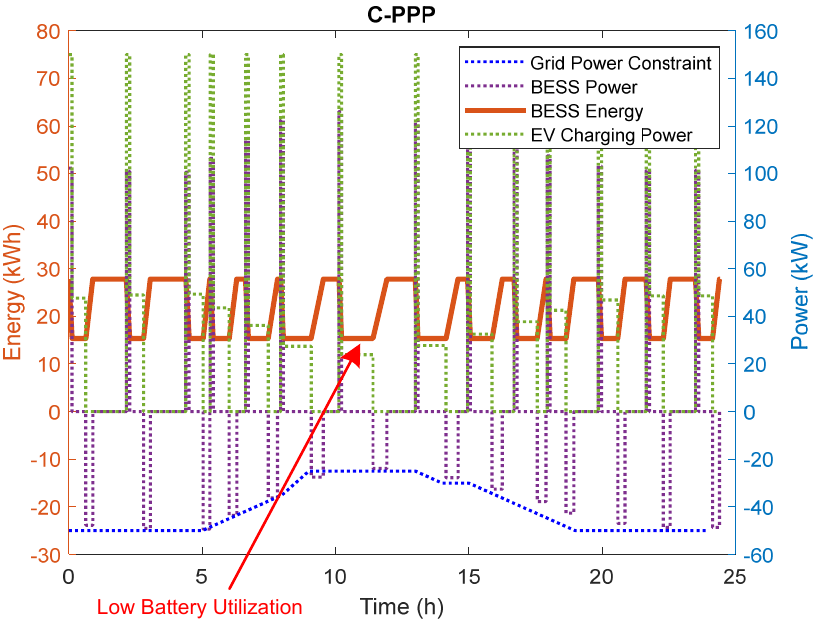} \label{fig:c-ppp-timeseries}}
    \caption{Single-Day Time Series for Three Actor Plaza.}
    \label{fig:time-series-3actor-plaza}
\end{figure}

A single-day time series illustrates the dynamic behavior of the EV charger, grid, and 2-BESS.  A comparison between a 2-BESS with LS-HiPPP and a 2-BESS with C-PPP is shown in Fig.~\ref{fig:time-series-3actor-plaza}. 

For LS-HiPPP, Fig.~\ref{fig:ls-hippp-timeseries} shows that nearly all the 2-BESS internal energy is utilized in comparison to C-PPP in Fig.~\ref{fig:c-ppp-timeseries}. C-PPP also results in a larger amount of curtailed charging from the smaller available 2-BESS energy outpout because of the worse battery energy utilization. Curtailed charging is indicated by the low power pedestals in the EV charging power trajectory. These pedestals correspond to using only the available grid power for EV charging.  C-PPP has 200\% more curtailed charging than LS-HiPPP. This indicates the EV charging time using C-PPP might be 2 times of the charging time using LS-HiPPP. 

This data suggests that 2-BESS can support fast charging. LS-HiPPP results in better performance with high utilization of the intrinsic battery energy and smaller curtailed charging pedestals.

\subsection{Ensemble Performance Over Single-Day Trajectory}
\begin{figure}[!t]
    \centering
    {\includegraphics[width=2.75 in]{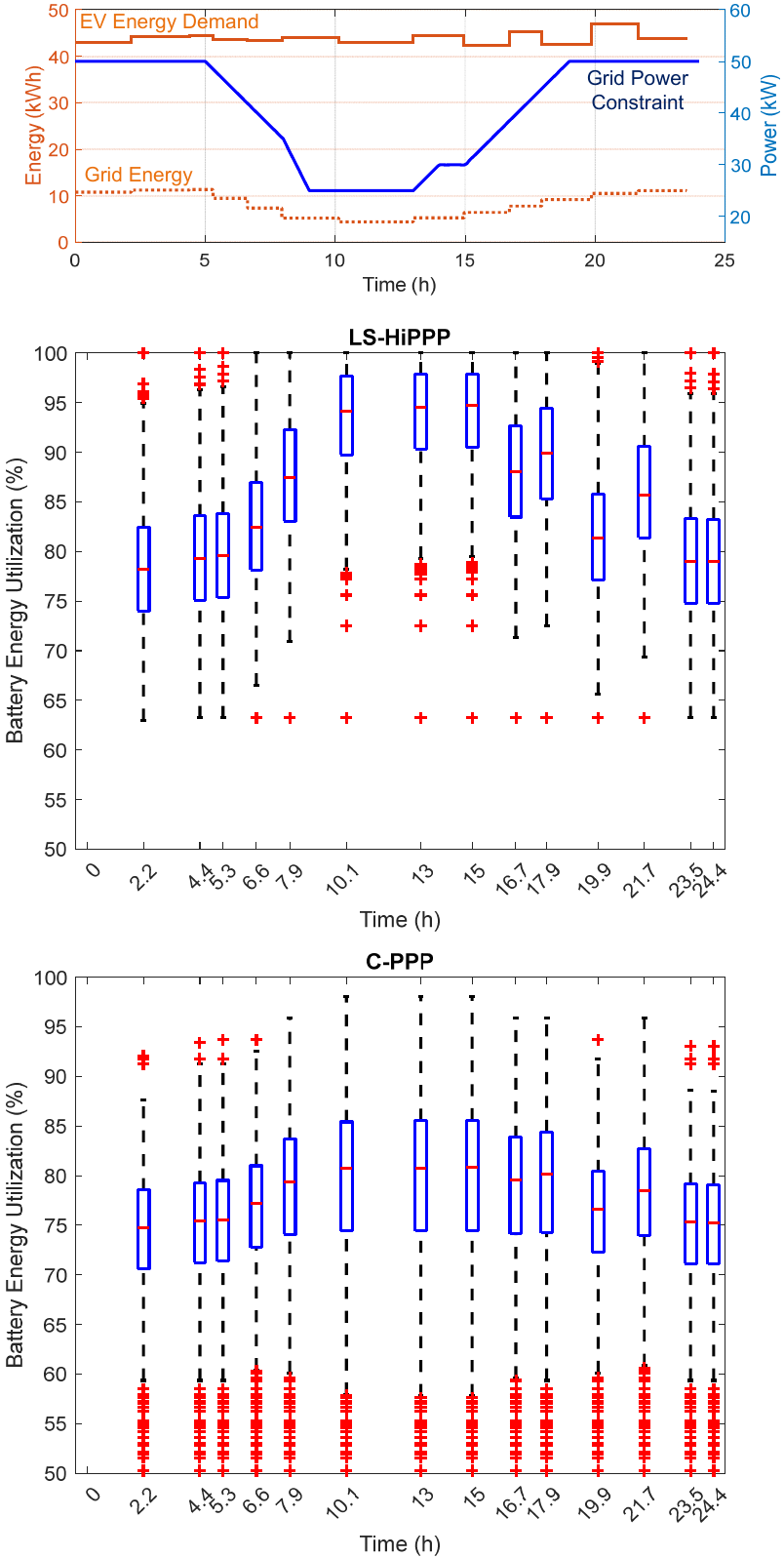} \label{fig:LS-HiPPP_CPPP_boxplot}}
    \caption{Box-Plot comparison of Battery Energy Utilization for a 25\% battery energy heterogeneity for a single-day trajectory exemplar.}
    \label{fig:boxplot}
\end{figure}

The statistical performance of LS-HiPPP and C-PPP over battery energy heterogeneity is evaluated over the same trajectory set in section~\ref{sec:single-day-3actor}.  Monte Carlo simulations were performed with individual battery energy capacity as the random variable.  The output random variable is battery energy utilization as shown in Fig.~\ref{fig:boxplot}. 

It can be observed that at high {\em grid-EV energy gap}, the mean battery energy utilization for LS-HiPPP is approximately 94.7\% versus 80.8\% for C-PPP.  It is worth noting that at low energy gap, the 2-BESS output may be low, which makes both LS-HiPPP and C-PPP underutilized as illustrated at 2.2\,h and 24.4\,h.

\subsection{Statistical Dispersion of Performance}
\begin{figure}[!t]
    \centering
    \subfloat[\label{fig:disper_mea_idr}IDR of 2-BESS Energy Output per EV Charge]{\includegraphics[width=2.5in]{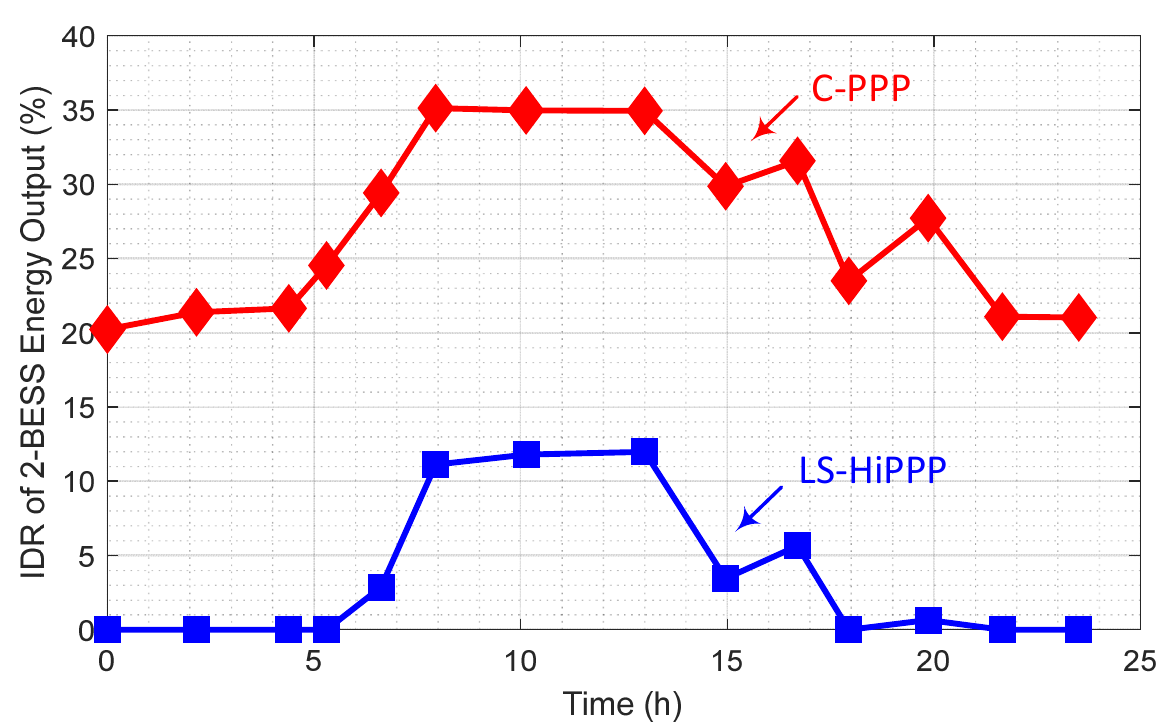}}
    \hfil
    \subfloat[\label{fig:disper_mea_3sigma} $3\sigma$ Spread of 2-BESS Energy Output per EV Charge]{\includegraphics[width=2.5in]{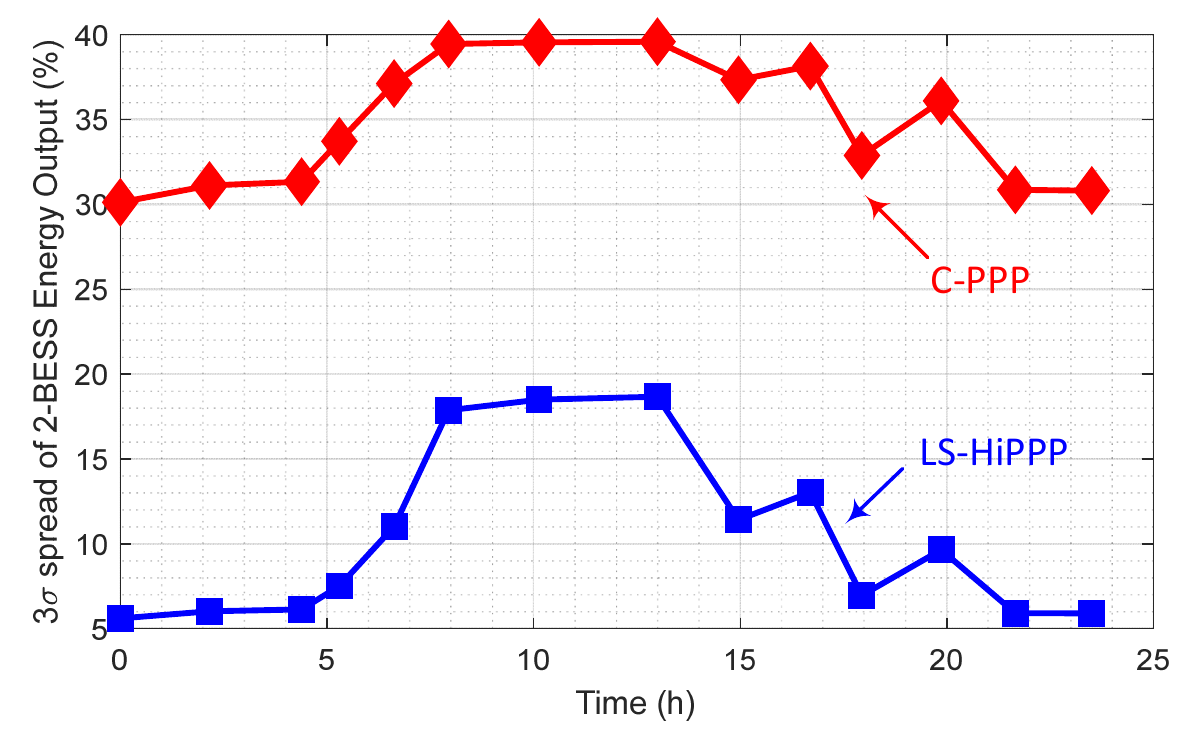}}
    \caption{\label{fig:disper_mea}Comparison of Statistical Dispersion of 2-BESS Energy Output.}
\end{figure}

A comparison of IDR if energy output among partial power processing architectures is shown in Fig.~\ref{fig:disper_mea_idr}. 
The IDR is significantly lower LS-HiPPP than that for C-PPP. Lower IDR at large grid-EV energy gap means higher consistency of performance when the 2-BESS is needed the most; from Fig.~\ref{fig:disper_mea_idr} at 10\,h, the IDR for LS-HiPPP is 10.4\% versus 32.7\% for C-PPP.  The $3\sigma$ spread is shown in Fig.~\ref{fig:disper_mea_3sigma} as a comparison.

\subsection{Resilience to Usage Uncertainty}

\begin{figure*}[!t]
    \centering
    \subfloat[LS-HiPPP]{\includegraphics[width=2.5in]{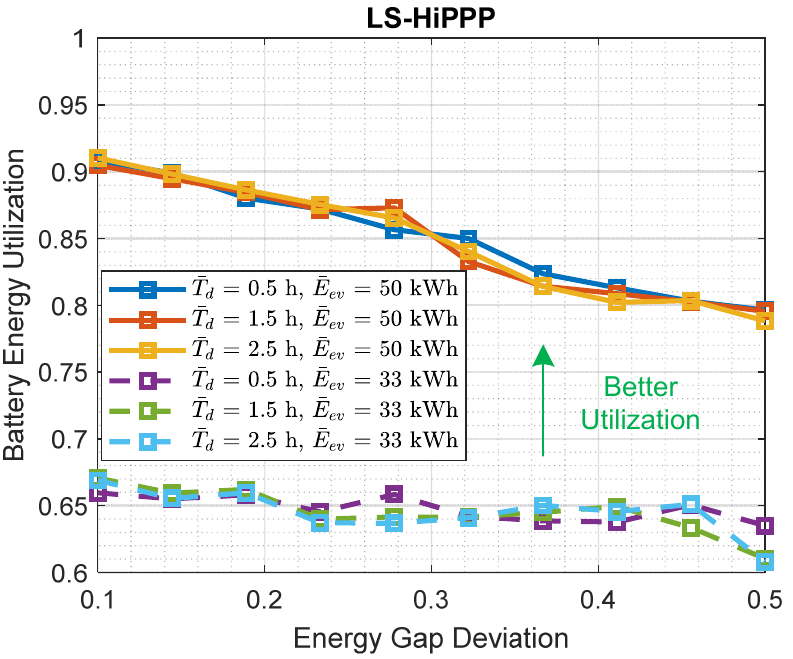} \label{fig:sto_var_gap_LSHiPPP}}
    \hfil
    \subfloat[C-PPP]{\includegraphics[width=2.6in]{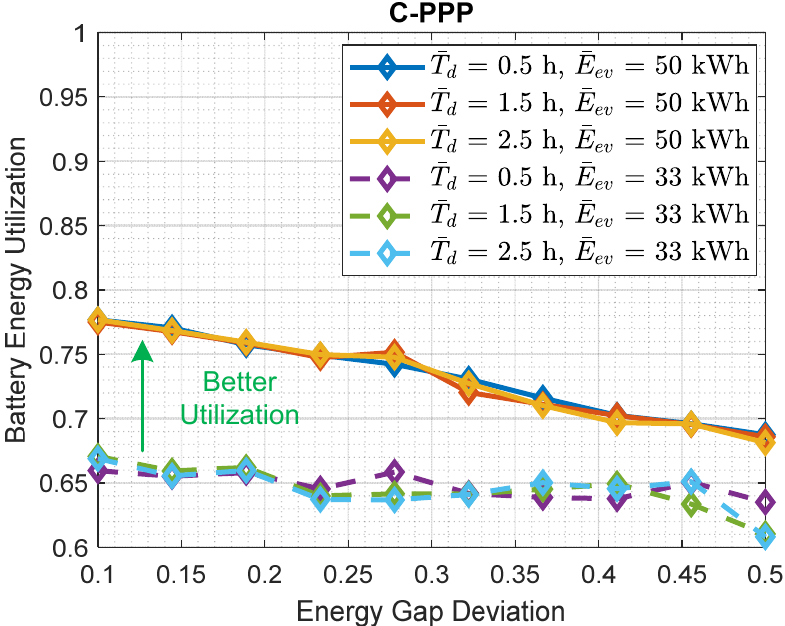} \label{fig:sto_var_gap_CPPP}}
    \caption{Comparison of the effect of stochastic deviation of grid-EV energy gap on battery energy utilization.}
    \label{fig:sto_var_gap}
\end{figure*}

By examining the effect of the stochastic deviation in the normalized grid-EV energy gap on expected battery energy utilization in Fig.~\ref{fig:sto_var_gap}, we can compare the resilience of LS-HiPPP and C-PPP to uncertainties in usage. 
$\bar{T}_d$ is the average standby time and $\bar{E}_{ev}$ is the average energy demand of EV charging.  It is worth noting that $\bar{T}_d$ does not affect the expected battery energy utilization.
At large energy gap deviation, e.g. 0.5, the utilization for LS-HiPPP is at 80\%, in comparison to that for C-PPP at 68\%.
For high $\bar{E}_{ev}$ (i.e., 50 kWh), the utilization decreases as a strong function of energy gap deviation. For low $\bar{E}_{ev}$ (i.e., 33\, kWh), the utilization is not significantly impacted by this deviation. 

\subsection{Curtailed Charging from Usage Uncertainty}
\begin{figure*}[!t]
    \centering
    \subfloat[LS-HiPPP]{\includegraphics[width=2.55in]{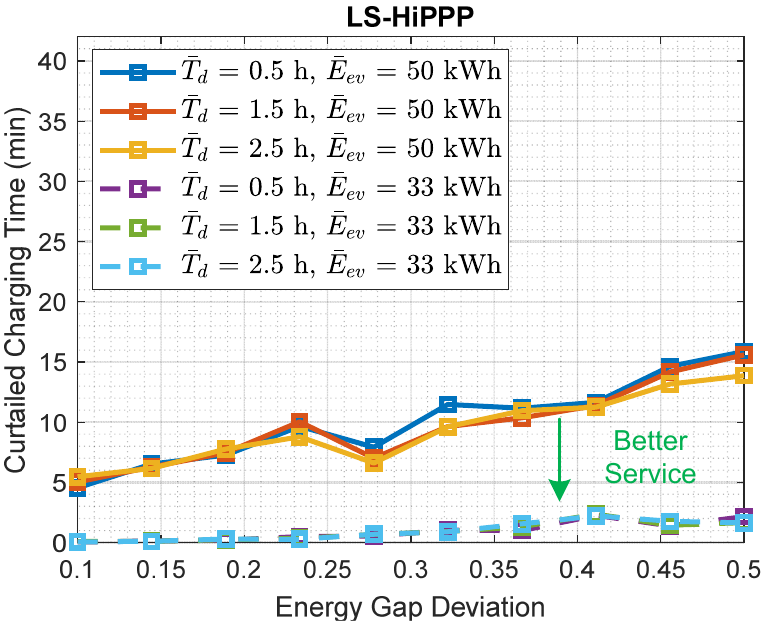} \label{fig:curt_charge_LSHiPPP}}
    \hfil
    \subfloat[C-PPP]{\includegraphics[width=2.5in]{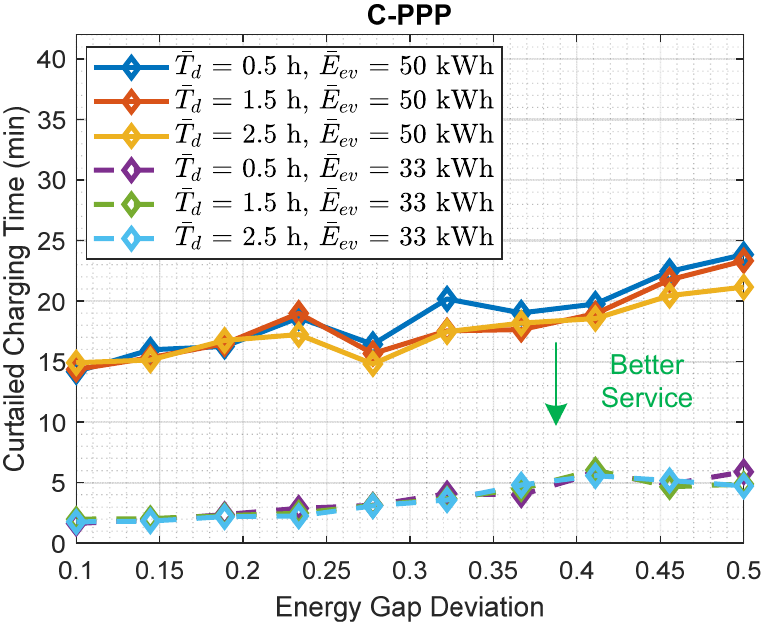} \label{fig:curt_charge_CPPP}}
    \caption{Comparison of the mean curtailed charging time for each EV.}
    \label{fig:curt_charge}
    \vspace{-10pt}
\end{figure*}

other When the stochastic deviation from expected usage is high, the 2-BESS is more likely to deplete in energy, resulting in curtailed EV charging. {\em Curtailed EV charging} occurs when the 2-BESS is depleted and EV charging power is reduced to only what the grid can instantaneously supply. Fig.~\ref{fig:curt_charge} shows that for a 50\% standard deviation from the mean value in the normalized grid-EV energy gap, the mean curtailed charging time for individual EVs is 67\% longer in C-PPP (25 minutes) than LS-HiPPP (15 minutes). This implies a higher quality of service to the EV charging customer when using LS-HiPPP by better mitigating the energy gap. In other words, LS-HiPPP enables a higher throughput of vehicles by maintaining a higher rate of EV charging. The charging time for a single EV using LS-HiPPP is significantly decreased compared to C-PPP.
\subsection{Derating and Confidence of Performance}
A derating factor can be inferred from the worst-case $3\sigma$ spread in Fig.~\ref{fig:disper_mea_3sigma}.  The derating factor for LS-HiPPP is significantly better at 84.3\% than C-PPP at 63.1\%.  This means that for a specified 2-BESS rating and confidence, LS-HiPPP requires a significantly smaller number of batteries and converters.
\subsection{Captured Value of the 2-BESS}
From the derating factor and battery energy utilization at the worst-case energy gap, the captured value for LS-HiPPP is 79.8\% versus 51.0\% for C-PPP.
\section{Conclusion} \label{sec:conclusion}
We investigated the performance of second-use battery energy storage systems (2-BESS) as an energy buffer for electrical vehicle chargers to interface with the grid. A new method for partial power processing (LS-HiPPP) within the 2-BESS was presented and compared with conventional partial power processing (C-PPP) and FPP (FPP) under: (i) supply heterogeneity, which imposed ensemble deviations in the battery energy capacity; (ii) time-varying grid capability; and (iii) stochastic EV charging demand.  LS-HiPPP shows significant performance improvements over C-PPP and FPP, with FPP falling well below the performance of C-PPP. LS-HiPPP outperforms C-PPP in the following metrics: battery energy utilization (94\% vs. 78\%), derating (84.3\% vs. 63.1\%), captured value (79.8\% vs. 51\%), resilience to usage uncertainty (80\% vs. 68\% in energy utilization at large grid-EV energy gap), and quality of service (15\,\text{minutes} vs. 25\,\text{minutes} in curtailed charging time), respectively.

Future work includes extending the model to more complex stochastic interactions between the grid, multiple EV chargers, and multiple 2-BESS units in a frequently used charging plaza, studying the tradeoff between the size of 2-BESS and the amount of EVs which the 2-BESS is capable of serving.
{
\bibliographystyle{ieeetr}
\bibliography{open_wo_revise.bib}
}

\end{document}